\documentclass[10pt,letterpaper,twocolumn]{article} 

\usepackage{ol2}
\usepackage[draft]{hyperref}
\usepackage{amsmath}
\begin{document}

\twocolumn[ 

\title{Two-Photon Dichroic Atomic Vapor Laser Lock Using Electromagnetically Induced Transparency and Absorption.}


\author{F. E. Becerra,$^{1,2}$ R. T. Willis,$^1$ S. L. Rolston,$^1$ and L. A. Orozco$^{1,*}$}

\address{
$^1$Joint Quantum Institute, Department of Physics, University of Maryland, and National Institute of Standards and Technology, College Park, MD 20742, U. S. A.\\
$^2$Departamento de F{\'i}sica, CINVESTAV, Apdo. Post. 14-740, 07000 M{\'e}xico, Distrito Federal, M{\'e}xico \\
$^*$Corresponding author: lorozco@umd.edu
}

\begin{abstract}
We demonstrate a technique to lock the frequency of a laser to a transition between two excited states in Rb vapor using a two-photon process in the presence of a weak magnetic field.  We use a ladder configuration from specific hyperfine sublevels of the $5S_{1/2}$, $5P_{3/2}$, and $5D_{5/2}$ levels. This atomic configuration can show Electromagnetically Induced Transparency and Absorption processes. The error signal comes from the difference in the transparency or absorption felt by the two orthogonal polarizations of the probe beam. A simplified model is in good quantitative agreement with the observed signals for the experimental parameters. We have used this technique to lock the frequency of the laser up to 1.5 GHz off atomic resonance.
\end{abstract}

 ] 

\noindent

\section{Introduction}
Methods for locking the frequency of a laser to an atomic transition between excited states are used in the study of spectroscopy \cite{perez07}, non-linear optics  \cite{becerra08, willis09, Sheng08}, and telecomunications \cite{boucher92, breton95}. Locking schemes using modulation of atomic resonances or light require demodulation to generate an error signal.
Methods with one laser use fluorescence detection after a degenerate two-photon excitation with light \cite{danielli00}, or atomic level modulation  \cite{baluschev00}.
Methods with two lasers modulate the frequency of one laser to generate an error signal for the other when it is scanned over resonance \cite{perez08, perez208, sheludko08}.

The frequency of the second laser can be locked to an excited state on resonance by the effects of optical pumping \cite{moon04}, two-photon absorption \cite{boucher92, breton95}, or through a non-linear process such as saturation, electromagnetically induced transparency (EIT) \cite{fleischhauer05}, and/or absorption (EIA) \cite{lezama99 , moon04b, yoon04,bellsc07, mullerdt08, abel08}.
Modulation-free methods generate an error signal from two signals with different nonlinear response from the same atomic reference
\cite{corwin98, robins02, wasik02, maric05}; in particular those methods using dichroism analyze the two polarization components of light that experience different nonlinearities in an atomic medium such as the Dichroic Atomic Vapor Laser Lock (DAVLL) \cite{corwin98}.

We present in this work a technique to stabilize the frequency of a laser to a transition between two excited states using a two-photon process in rubidium vapor.  It is based on the principles of a  DAVLL lock. The ladder configuration of the two-photon transition allows us to use EIT and an increase in absorption due to two-photon absorption (TPIA) \cite{gea95} to obtain a narrow response with tuning range greater than 1 GHz around the atomic transition. Our scheme is modulation-free. The method profits from the precision of a Doppler-free two-photon process and the tunability of the DAVLL technique, requiring low power for its implementation.

The organization of the paper is as follows: Sec. \ref{Experiment} presents the experimental configuration, Sec. \ref{Theory} has the theoretical treatment. Sec. \ref{Results} contains the experimental results with their analysis on Sec. \ref{Analysis}. Sec. \ref{Conclusions} summarizes the conclusions.

\section{Experiment}\label{Experiment}

\begin{figure}
\centering
\includegraphics[width=7.5cm]{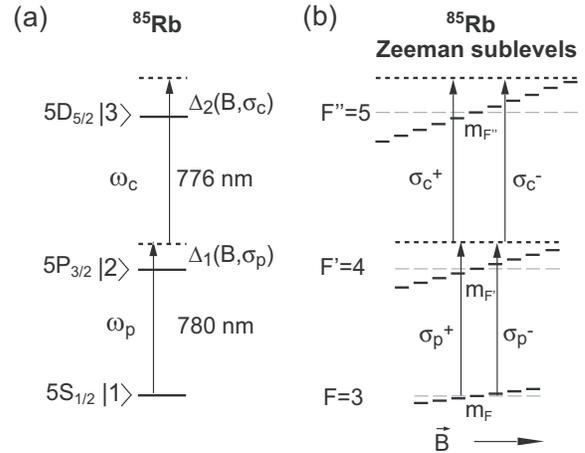}
\caption{\label{fig:Level} Schematic of the atomic levels used and the experimental layout. (\textbf{a}) three-level atomic model and energy levels in Rb, (\textbf{b}) Zeeman magnetic sublevels of the $5S_{1/2},F\mathord{=}3$, $5P_{3/2},F'\mathord{=}4$, and $5D_{5/2},F''\mathord{=}5$ hyperfine levels in $^{85}$Rb in the presence of a small magnetic field $\vec{B}$ (The splitting of the Zeeman sublevels is not to scale).}
\end{figure}

Figure \ref{fig:Level}(a) shows a simplified model for the ladder configuration in Rb between the $5S_{1/2}$, $5P_{3/2}$, and $5D_{5/2}$ levels.
Fig. \ref{fig:Level}(b) shows a more realistic configuration taking into account the specific polarization and hyperfine levels that we use as well as the Zeeman splitting in the presence of a magnetic field.
We select specific hyperfine states and, given the small magnetic fields present, do not resolve the different magnetic Zeeman sublevels.
This configuration shows EIT and TPIA of a probe beam resonant with the $|1\rangle\mathord{\rightarrow}|2\rangle$ transition, in the presence of a coupling beam resonant with the $|2\rangle\mathord{\rightarrow}|3\rangle$ transition propagating in opposite direction when they satisfy two-photon resonance \cite{gea95}.
We use a rubidium cell at room temperature with natural abundance of $^{85}$Rb and $^{87}$Rb isotopes as our atomic system, and although we describe the work done with $^{85}$Rb, the same results hold for $^{87}$Rb.

\begin{figure}
\centering
\includegraphics[width=6.5cm]{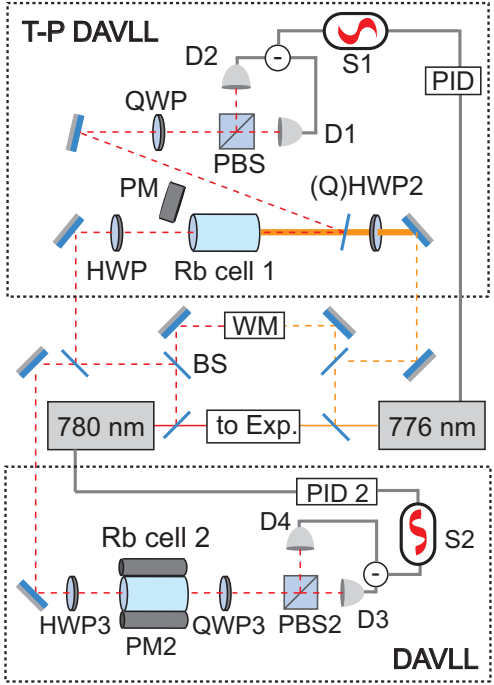}
\caption{\label{fig:Exp} (Color online) Experimental setup for the two-photon lock; Two-Photon DAVLL (Top) and DAVLL (Bottom): (P)BS (polarizing) beam splitters; PM, permanent magnets; D, detectors; (Q)HWP (quarter) half wave plates; WM, wavemeter; PID, feedback controllers; S1, S2, subtraction signals.}
\end{figure}

Fig. \ref{fig:Exp} shows the experimental configuration:
 a weak probe beam (780 nm) propagates along the axis of a rubidium vapor cell (Rb cell 1) with linear polarization (see Fig. \ref{fig:Exp} top). The coupling beam (776 nm) propagates in the opposite direction overlapping with the probe beam in the cell. The probe beam experiences EIT or TPIA, depending on the intermediate state detuning, when the two lasers satisfy the two-photon resonance. We detect the two circular polarization components of the probe laser, with a quarter wave plate (QWP) and a polarizing beam splitter (PBS), with the two photodetectors D1 and D2.

A permanent magnet (PM) produces a field of $B\simeq2$ G pointing along the axial direction of the cell to shift the EIT and TPIA for the  different probe polarizations by the Zeeman effect, and produce a dispersion-like subtraction signal S1.
The diode laser probe (780 nm) with a linewidth of 10 MHz (measured using a homodyne technique \cite{okoshi80}) propagates along the axial direction of a 7-cm rubidium cell at room temperature. It has linear polarization, a power of 400 $\mu$W, and a diameter $1\mathord{/}e^{2}$ of 0.5 mm. A Titanium:Sapphire coupling laser with a linewidth of 100 kHz,  propagates in the opposite direction overlapping the probe beam along the cell with linear polarization, a power of 4 mW, and a diameter of 1 mm. We detect the different circular polarization components of the probe beam, defined by the quantization axis along the beam propagation direction,
in the detectors D1 and D2 with 780 bandpass filters using a QWP and a PBS (See Fig \ref{fig:Exp}).

\section{Theory}\label{Theory}
One way to describe the effects on a weak probe beam due to the atomic medium in the presence of the coupling beam in the ladder configuration is solving the density matrix equations under some approximations \cite{gea95}. When we consider the atomic system shown in Fig. \ref{fig:Level} (a), the laser detunings are $\Delta_{1}\mathord{=}\omega_{p}\mathord{-}\omega_{21}$ and $\Delta_{2}\mathord{=}\omega_{c}\mathord{-}\omega_{32}$, with $\omega_{21}$ and $\omega_{32}$ the atomic transition frequencies of $|1\rangle\mathord{\rightarrow}|2\rangle$ and $|2\rangle\mathord{\rightarrow}|3\rangle$ respectively, and $\omega_{p}$ and $\omega_{c}$ the probe and coupling laser frequencies respectively. The total decay rates are $\Gamma_{ij}\mathord{=}(\Gamma_{i}\mathord{+}\Gamma_{j})/2$; where $\Gamma_{i}$ is the natural decay rate of the level $|i\rangle$, and $\mu_{ij}$ is the dipole moment matrix element of the transition $|j\rangle\mathord{\rightarrow}|i\rangle$.
The atomic polarization $P(\omega_{p})$ at the probe field frequency is proportional to the density matrix element $\rho_{21}$ that takes a simple form under some approximations:
assuming weak probe field $E_{p}$, we keep the lowest order in $E_{p}$; solving the density matrix equations in steady state; and, since the probe field is weak, the populations in levels $|2\rangle$ and $|3\rangle$ are negligible in comparison with the population in level $|1\rangle$ that is approximately unity \cite{gea95}:
\begin{equation}
\rho_{21}\mathord{\simeq}\mathord{-}\frac{ig_{21}}{\displaystyle  \gamma_{21}\mathord{-}i\Delta_{1}\mathord{+}\frac{\Omega_{c}^2/4}{\gamma_{31}\mathord{-}i(\Delta_{1}\mathord{+}\Delta_{2})}}E_{p},
\label{ro21}
\end{equation}
where $2g_{ij}\mathord{=}\mu_{ij}\mathord{/}\hbar$ and $\Omega_{c}\mathord{=}2g_{32}E_{c}$ is the Rabi frequency of the coupling field $E_{c}$. The susceptibility at the probe frequency is then proportional to the density matrix element $\rho_{21}$ shown in Eq. (\ref{ro21}) whose real and imaginary parts describe the dispersive and absorptive properties of the medium respectively.

The motion of the atoms causes Doppler broadening in the atomic medium. The counterpropagating beam configuration of the probe and coupling beams  reduces this effect because an atom with velocity $v$ will have a two-photon detuning $\Delta_{1}\mathord{+}\Delta_{2}\mathord{+}(\omega_{p}\mathord{-}\omega_{c})v/c$ in the ladder configuration. When the probe and coupling frequencies are equal the process is Doppler free.
The susceptibility at the probe frequency of the Doppler medium that describes the absorption and dispersion experienced by a weak probe beam in the cell with a counterpropagating coupling beam in the ladder configuration shown in Fig. \ref{fig:Level} (a) \cite{gea95} is
\begin{equation}
\chi\mathord{=}\int\frac{(4i\hbar |g_{21}|^{2}/\epsilon_{0})~N(v)}{\displaystyle  \gamma_{21}\mathord{-}i\Delta_{1}\mathord{-}i\frac{\omega_{p}}{c}v\mathord{+}\frac{\Omega_{c}^2/4}{\gamma_{31}\mathord{-}i\Delta_{2\gamma}\mathord{-}i\Delta\omega_{L}v/c}}dv,
\label{chi}
\end{equation}
where $\Delta_{2\gamma}\mathord{=}\Delta_{1}\mathord{+}\Delta_{2}$ is the two-photon detuning, and the laser frequency mismatch is $\Delta\omega_{L}\mathord{=}\omega_{p}\mathord{-}\omega_{c}$. The velocity ($v$) distribution of the atoms in the cell is $N(v)\mathord{=}(N_{0}/u\sqrt{\pi})\exp\left(-v^2/u^2\right)$ with the root-mean-square velocity $u/\sqrt{2}$, and $c$ is the speed of light in vacuum. The effective linewidths are $\gamma_{21}\mathord{=}\Gamma_{21}\mathord{+}\gamma_{p}$ and $\gamma_{31}\mathord{=}\Gamma_{31}\mathord{+}\gamma_{p}\mathord{+}\gamma_{c}$ where $\gamma_{p}$ and $\gamma_{c}$ are the laser linewidths.

\begin{figure}
\centering
\includegraphics[width=8.cm]{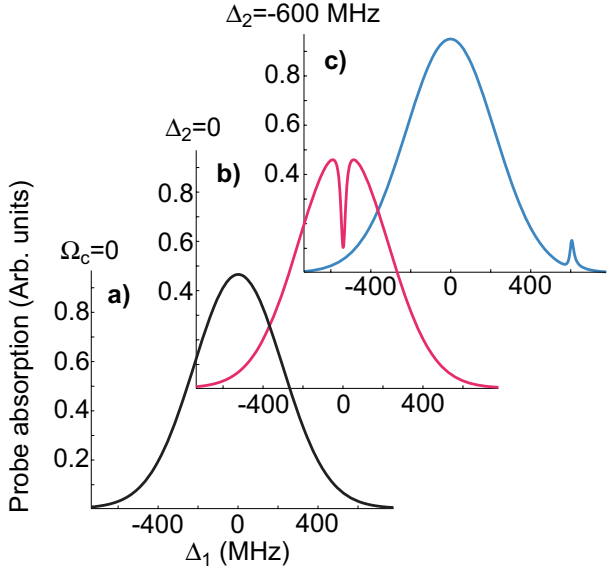}
\caption{\label{fig:EIT_Example} (Color online)  Probe absorption with and without coupling field of $\Omega_{c}\mathord{=}80$MHz from Eq. (\ref{chi}). (\textbf{a}) in the absence of coupling field the probe shows a Doppler absorption profile. (\textbf{b}) with the coupling field on resonance the probe experiences EIT.  (\textbf{c}) with the coupling field off resonance the probe experiences an increase in absorption due to two-photon absorption (TPIA) (The parameters used for these plots are similar to our experimental parameters).}
\end{figure}

Figure \ref{fig:EIT_Example} shows the modification the probe absorption for different coupling powers and detunings using Eq. (\ref{chi}).
The absorption of the probe beam shows a Doppler broadened profile around the one-photon resonance when the coupling beam is absent (see Fig. \ref{fig:EIT_Example}(a)). The presence of the coupling beam induces a two-photon process that affects the absorption of the probe beam, EIT or TPIA, as a function of the detunings of both beams from the respective atomic transitions (see Fig. \ref{fig:EIT_Example}(b,c)).
The two-photon resonance condition, $\Delta_{1}\mathord{+}\Delta_{2}\mathord{=}0$, determines the location of the EIT window as well as the TPIA of the probe beam \cite{gea95}. A small modification on the energy levels will displace the EIT or TPIA experienced by the probe beam. We use a small magnetic field to shift the levels for a particular probe polarization by the Zeeman Effect.

We model the system by assuming that the two-photon processes for all the Zeeman sublevels are independent, since the transitions are coupled by orthogonal polarizations, and the population remains mostly in the ground state for a weak probe \cite{gea95}. We assume that the population is distributed homogeneously in the Zeeman ground states.
We use Eq. (\ref{chi}) for each possible two-photon transition among Zeeman magnetic sublevels of the ground, $m_{F}$, intermediate, $m_{F'}$, and upper state, $m_{F''}$, with the respective Zeeman shifted atomic transitions (see Fig. \ref{fig:Level}(b)) for the different circular probe and coupling polarizations. The magnetic field ($B$) modifies the probe and coupling detunings, $\Delta_{1}(m_{F},m_{F'})$ and $\Delta_{2}(m_{F'},m_{F''})$, depending on the Land{\'e} factors and the particular Zeeman sublevels:
\begin{eqnarray}
\Delta_{1}(m_{F},m_{F'})\mathord{=}\Delta_{1}^{0}\mathord{+}(m_{F}g_{F}\mathord{-}m_{F'}g_{F'})\mu_{B}B\nonumber\\
\Delta_{2}(m_{F''},m_{F''})\mathord{=}\Delta_{2}^{0}\mathord{+}(m_{F'}g_{F'}\mathord{-}m_{F''}g_{F''})\mu_{B}B,
\label{modified detunings}
\end{eqnarray}
where $g_{F}\mathord{=}0.2708$, $g_{F'}\mathord{=}0.6182$ and $g_{F''}\mathord{=}0.7631$ for the specific hyperfine transitions in our experiment, $\mu_{B}\mathord{\simeq}1.4$ MHz/G is the Bohr magneton and $\Delta_{i}^{0}$ are the detunings of the lasers from the zero B-field atomic resonances.
The dipole matrix element for the coupling $g_{32}(m_{F'},m_{F''})$ and the probe $g_{21}(m_{F},m_{F'})$ characterize each transition strength:
\begin{eqnarray}
g_{21}(m_{F},m_{F'})&\mathord{=}&g_{21}^{0}(F,F')A_{1}(F,F')\nonumber \\
&\times&(-1)^{m_{F}}
\left(
  \begin{array}{ccc}
    F' & 1 & F \\
    m_{F'} & q_{1} & -m_{F} \\
  \end{array}
\right)
\nonumber\\
g_{32}(m_{F},m_{F'})&\mathord{=}&g_{32}^{0}(F',F'')A_{2}(F',F'')\nonumber \\
&\times&(-1)^{m_{F'}}\left(
  \begin{array}{ccc}
    F'' & 1 & F' \\
    m_{F''} & q_{2} & -m_{F'} \\
  \end{array}
\right)
\label{relative dipole elements}
\end{eqnarray}
with $m_{F}\mathord{=}m_{F'}\mathord{+}q_{1}$ and $m_{F'}\mathord{=}m_{F''}\mathord{+}q_{2}$. The diploe matrix elements $g_{21}^{0}(F,F')$ and $g_{32}^{0}(F',F'')$ are common factors as well as $A_{1}(F,F')$ and $A_{2}(F',F'')$ and they are not discussed in the present work. The quantities in round brackets are $3\mathord{-}J$ symbols and describe the relative strength of the dipole elements for different Zeeman sublevels in a dipole transition. Each dipole matrix element for transitions with defined Zeeman sublevels modifies the relative strength of each two-photon process depending on the polarizations of the probe and coupling beams. We calculate the total susceptibility for the two circular probe polarizations $\sigma^{+}$ and $\sigma^{-}$ using Eq. (\ref{chi}) with the modified detunings in Eq. (\ref{modified detunings}) and relative dipole matrix elements in Eq. (\ref{relative dipole elements}) for each two-photon process, and adding all the respective contributions for the different probe polarizations.
The Two-Photon DAVLL (T-P DAVLL) signal S1 is the subtraction of these atomic responses of the two components of the probe light associated with the absorption:
\begin{eqnarray}
&S1\mathord{=}\mathbf{Im}(\chi^{\sigma^{+}}\mathord{-}\chi^{\sigma^{-}}),\nonumber \\ &\chi^{\sigma^{+\mathord{(-)}}}\mathord{=}\displaystyle\sum_{m_{F,F',F''}}^{\sigma^{+\mathord{(-)}}}\chi(m_{F},m_{F'},m_{F''})
\label{S1}
\end{eqnarray}
with $m_{F'}\mathord{=}m_{F}\mathord{+}1$ and $m_{F'}\mathord{=}m_{F}\mathord{-}1$ for $\sigma^{+}$ and $\sigma^{-}$ probe polarization respectively.


\section{Results}\label{Results}

\begin{figure}
\centering
\includegraphics[width=7.2cm]{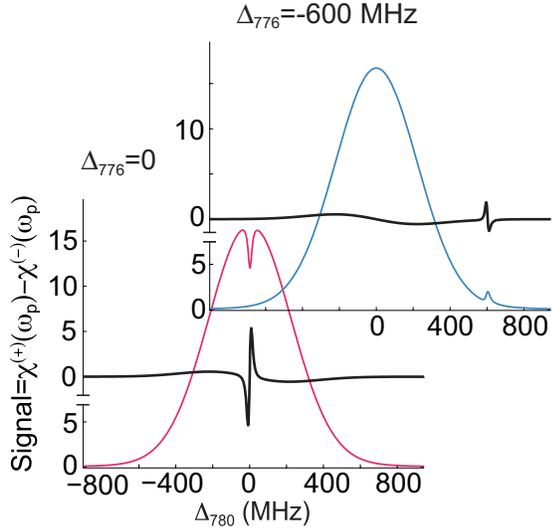}
\caption{\label{fig.EIT-EIA} (Color online)  Theoretical prediction of the S1 signal form Eq. (\ref{S1}) as a function of probe detuning for different coupling detunings. Theoretical T-P DAVLL signal$\times$5 in black: front plane, from EIT  in dark gray (red); back plane, from \text{TPIA}  in light gray (blue).}
\end{figure}

\begin{figure}
\centering
\includegraphics[width=8.2cm]{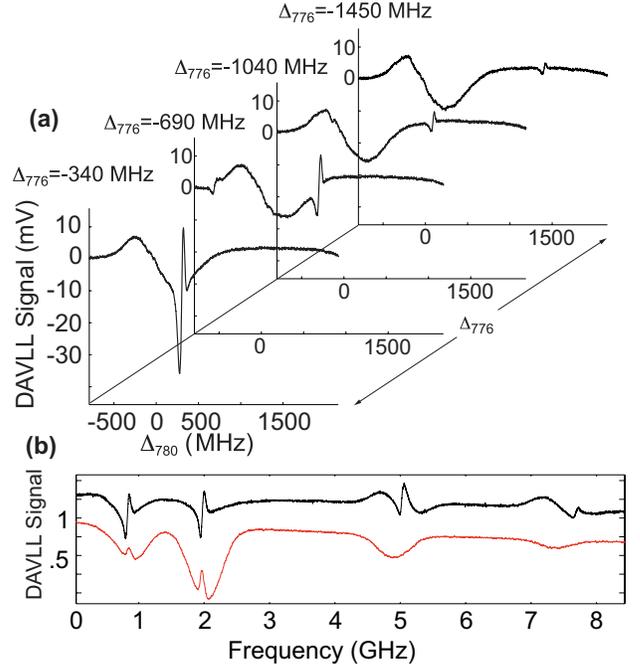}
\caption{\label{fig:Scan780DAVLL} (Color online)  Experimental Two-Photon (T-P DAVLL) signal as a function of probe (780 nm) frequency detuning for different coupling (776 nm) detunings. (\textbf{a}) experimental T-P DAVLL Signal
for 780 nm laser  scanning on the transition $5S_{1/2},F\mathord{=}3\mathord{\rightarrow}5P_{3/2},F'\mathord{=}4$ in $^{85}$Rb. (\textbf{b}): black line, T-P DAVLL signal for different levels in Rb for coupling laser on resonance; gray (red) line, absorption signal in detector D1.}
\end{figure}

Figure \ref{fig.EIT-EIA} shows the theoretical  T-P DAVLL signal S1$\times5$ from Eq. (\ref{S1}) as a function of probe frequency for different coupling frequencies for the experimental parameters of $B\mathord{=}2$ G, temperature $21^{\circ}$C, $\gamma_{21}\mathord{=}14$  MHz and $\gamma_{31}\mathord{=}12.5$ MHz, and coupling Rabi frequency $\Omega_{c}\mathord{=}80$ MHz.
We show two different coupling frequencies that produce EIT (front plane) and TPIA (back plane) of the probe beam of a given circular polarization, the T-P DAVLL can be produced form these two different processes.
Fig. \ref{fig:Scan780DAVLL} shows experimental result for a given probe and coupling laser intensities (see Fig. \ref{fig:Exp}), when the probe and coupling lasers are resonant with the transitions $5S_{1/2},F\mathord{=}3\mathord{\rightarrow}5P_{3/2},F'\mathord{=}4$ and $5P_{3/2},F'\mathord{=}4\mathord{\rightarrow}5D_{5/2},F''\mathord{=}5$ in $^{85}$Rb. This signal has a dispersion-like frequency dependence, adequate for frequency stabilization of the coupling light resonant with a two-photon process. When the detuning of the coupling laser is smaller than the Doppler width, the probe experiences EIT, and the T-P DAVLL S1 is generated from two EIT processes (See Fig. \ref{fig:Scan780DAVLL}(b), note the probe frequency scan over $^{85}$Rb and $^{87}$Rb). If the detuning of the coupling laser exceeds the Doppler width, the probe laser experiences TPIA, and it generates a dispersion-like signal that allows us to stabilize the frequency of the coupling laser over a range of 2 GHz detuned from resonance as shown in Fig \ref{fig:Scan780DAVLL}(a). The sign of the signal with TPIA has to be compensated appropriately by electronic means to lock as shown in this figure.

\begin{figure}
\centering
\includegraphics[width=6cm]{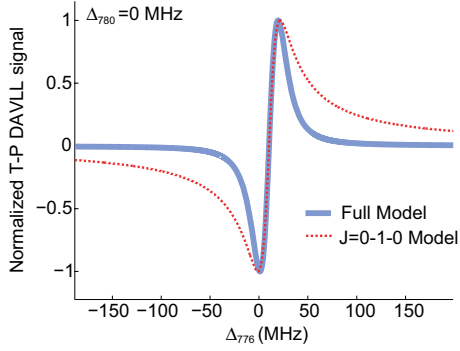}
\caption{\label{Compare-Models}
(Color online) Expected T-P DAVLL signal for frequency stabilization as a function of coupling laser detuning for probe laser locked on resonance
for two different models: thick line, Full model from Eq. (\ref{S1}); dashed line, $J\mathord{=}0\mathord{-}1\mathord{-}0$ with Zeeman sublevels $m_{F}\mathord{=}0$, $m_{F'}\mathord{=\pm}1$ and $m_{F''}\mathord{=}0$ (both signals normalized for comparison).}
\end{figure}

Figure \ref{Compare-Models} shows the expected T-P DAVLL signal as a function of coupling laser detuning for the probe resonant with the atomic transition for two different models: full model form Eq. (\ref{S1}) that takes in account all the Zeeman sublevels $m_{F}$, $m_{F'}$ and $m_{F''}$, and a simplified model $J\mathord{=}0\mathord{-}1\mathord{-}0$ with Zeeman sublevels $m_{F}\mathord{=}0$, $m_{F'}\mathord{=\pm}1$ and $m_{F''}\mathord{=}0$ (both signals are normalized for comparison). The $J\mathord{=}0\mathord{-}1\mathord{-}0$ model forms two ladder EIT configurations with opposite probe and pump circular polarizations with the assumptions described above, and shows the same slope around resonance of the full model can show (given proper normalization). The lineshapes differ because all the other contributions from different paths among all Zeeman sublevels and their relative strengths. The main effect this has is the reduction of the frequency range where the feedback signal is not zero and has the adequate sign for feedback control. Taking all the m sublevels into consideration reduces this effective frequency range.

The frequency range where the error signal is monotonic and crosses zero, capture range, increases linearly with the coupling intensity. The slope of the error signal decreases with the coupling Rabi frequency until for sufficiently large coupling intensity gets distorted.

\section{Analysis}\label{Analysis}
\begin{figure}
\centering
\includegraphics[width=8.5cm]{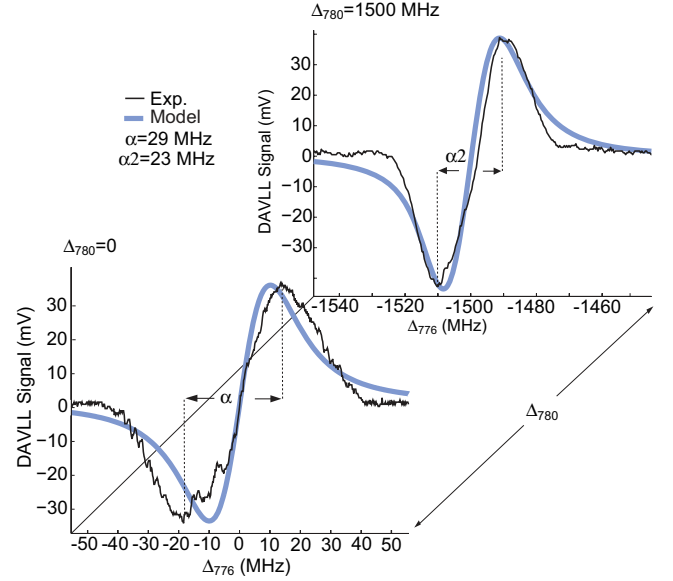}
\caption{\label{fig:T-P-DAVLL-SIGNAL}
(Color online) Single shot of experimental (thin line) and modeled (thick line) T-P DAVLL signal for frequency stabilization as a function of 776 nm laser detuning for 780 nm laser locked on resonance (front plane) and 1.5 GHz off resonance (back plane).}
\end{figure}

\begin{figure}
\centering
\includegraphics[width=8.5cm]{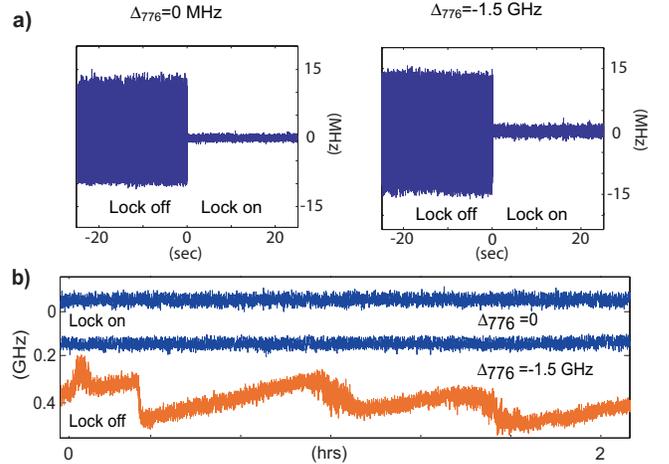}
\caption{\label{fig:LOCKSTABILITY}
(Color online) Lock stability. \textbf{a}) in-loop lock noise reduction of the coupling laser for $\Delta_{776}=0$ and $\Delta_{776}=-1.5$GHz. \textbf{b}) absolute frequency of the coupling laser over 2 hours measured with a wave meter. Dark (blue in color) lines, laser locked;  light gray (orange in color) line, laser unlocked.}
\end{figure}

We achieve the frequency stabilization of the coupling laser by locking the probe laser frequency with a traditional DAVLL setup  \cite{corwin98} as shown in Fig. \ref{fig:Exp} in the bottom. This setup is protected from temperature fluctuations with thermal insulation. It provides enough tunability and stability for the two-photon resonance frequency stabilization of the coupling laser. We scan the coupling laser on the two-photon resonance and detect the signal S1.
Fig. \ref{fig:T-P-DAVLL-SIGNAL} shows single-shot experimental (thin line) and theoretical
expected (thick line) T-P DAVLL S1 signals as a function of coupling laser detuning for the probe laser locked on resonance with a coupling power of 5 mW on the front plane,  and 1.5 GHz detuned from resonance with a coupling power of 10 mW on the back plane (coupling
Rabi frequency $\Omega_{c}\mathord{=}80$ MHz, with the amplitude of the theory a free parameter).
We use this signal and a PID (Proportional, Integral and Differential) feedback control to lock the frequency of the coupling laser.
Fig. \ref{fig:LOCKSTABILITY}(a) shows the in-loop noise frequency reduction of the coupling laser for the two detunings, as seen in the error signal, with a Gaussian frequency distribution with $\sigma\mathord{=}360$ kHz and $\sigma\mathord{=}1.1$ MHz  for the on and off resonance case respectively with the laser locked.
Homodyne measurements of the locked coupling laser linewidth show a Gaussian distribution with a $\sigma\mathord{=}550$ kHz.

Tests of the long term frequency stability with a wave meter (4 Hz sampling rate,
repeatability of  $5\mathord{\times}10^{-8}$  over 10 minutes),  show Gaussian fractional distributions  of fluctuations with $\sigma\mathord{=}4.19\mathord{\times}10^{-8}$ (on resonance) and fractional fluctuations of $\sigma\mathord{=}4.62\mathord{\times}10^{-8}$ (off resonance) for a run of two hours. Fig. \ref{fig:LOCKSTABILITY}(b) shows the reading of the wave meter over two hours: dark gray (blue) lines, laser locked; light gray (orange) lines, laser unlocked.
The wave meter gives a digital output with a discrete resolution ($\approx$ 20 MHz).

\section{Conclusions}\label{Conclusions}
 We have developed a modulation-free scheme to stabilize the frequency of a laser to a transition between two excited states resonant with a two-photon process with a tuning range over 2 GHz assisted with EIT and TPIA in a ladder configuration. Our scheme combines the Doppler-free properties of a two-photon process and the tunability of a traditional DAVLL technique.\\

This work was supported by NSF and CONACYT.


\begin{thebibliography}{10}
\newcommand{\enquote}[1]{``#1''}

\bibitem{perez07}
A. P\'{e}rez Galv\'{a}n, Y.~Zhao, L.~A. Orozco, E.~G\'{o}mez, A.~D. Lange, F.~Baumer,
  and G.~D. Sprouse, \enquote{Comparison of hyperfine anomalies in the 5S$_{1/2}$
  and 6S$_{1/2}$ levels of $^{85}$Rb and $^{87}$Rb,} Phys. Lett. B \textbf{655}, 114-118 (2007).

\bibitem{becerra08}
F.~E. Becerra, R.~T. Willis, S.~L. Rolston, and L.~A. Orozco,
  \enquote{Nondegenerate four-wave mixing in rubidium vapor: The diamond
  configuration,} Phys. Rev. A \textbf{78}, 013834 (2008).

\bibitem{willis09}
R.~T. Willis, F.~E. Becerra, L.~A. Orozco, and S.~L. Rolston,
  \enquote{Four-wave mixing in the diamond configuration in an atomic vapor,} Phys. Rev. A \textbf{79}, 033814 (2009).

\bibitem{Sheng08}
D. Sheng, A. P\'{e}rez Galv\'{a}n, and L. A. Orozco
"Lifetime measurements of the 5d states of rubidium"
Phys. Rev. A \textbf{78}, 062506 (2008).

\bibitem{boucher92}
R.~Boucher, M.~Breton, N.~Cyr, and M.~Tetu, \enquote{Dither-free absolute
  frequency locking of a 1.3 $\mu$m dfb laser on $^{87}$Rb,} IEEE Photon. Technol. Lett.
  \textbf{4}, 327-329 (1992).

\bibitem{breton95}
M.~Breton, P.~Tremblay, C.~Julien, N.~Cyr, M.~Tetu, and C.~Latrasse,
  \enquote{Optically pumped rubidium as a frequency standard at 196 THz,} IEEE
  Trans. Instrum. Meas. \textbf{44}, 162-165 (1995).


\bibitem{danielli00}
A.~Danielli, P.~Rusian, A.~Arie, M.~H. Chou, and M.~M. Fejer,
  \enquote{Frequency stabilization of a frequency-doubled 1556-nm source to the
  5S$_{1/2}$ - 5D$_{5/2}$ two-photon transitions of rubidium,} Opt. Lett. \textbf{12},
  905-907 (2000).

\bibitem{baluschev00}
S.~Baluschev, N.~Friedman, L.~Khaykovich, D.~Carasso, B.~Johns, and
  N.~Davidson, \enquote{Tunable and frequency-stabilized diode laser with a
  Doppler-free two-photon Zeeman lock,} Appl. Opt. \textbf{39}, 4970-4974 (2000).

\bibitem{perez08}
A. P\'{e}rez Galv\'{a}n, Y.~Zhao, and L.~A. Orozco, \enquote{Measurement of the
  hyperfine splitting of the 6S$_{1/2}$ level in rubidium,} Phys. Rev. A
  \textbf{78}, 012502 (2008).

\bibitem{perez208}
A. P\'{e}rez Galv\'{a}n, D. Sheng, L. A. Orozco, and Y. Zhao, "Two Color Modulation Transfer Spectroscopy," e-print
arXiv:0812.1386v1, Can. J. Phys. to be published.

\bibitem{sheludko08}
David V. Sheludko, Simon C. Bell, Russell Anderson, Christoph S. Hofmann, Edgar J. D. Vredenbregt, and Robert E. Scholten, \enquote{State-selective imaging of cold atoms,} Phys. Rev. A \textbf{77}, 033401 (2008).

\bibitem{moon04}
H.~S. Moon, W.~K. Lee, L.~Lee, and J.~B. Kim, \enquote{Double resonance optical
  pumping spectrum and its application for frequency stabilization of a laser
  diode,} Appl. Phys. Lett. \textbf{85}, 3965 (2004).

\bibitem{fleischhauer05}
Michael Fleischhauer, Atac Imamoglu, and Jonathan P. Marangos,
  \enquote{Electromagnetically induced transparency: Optics in coherent media,} Rev. Mod. Phys.
 \textbf{77}, 633-673 (2005).

\bibitem{lezama99}
A. Lezama, S. Barreiro, and A. M. Akulshin, \enquote{Electromagnetically induced absorption,}
Phys. Rev. A \textbf{59}, 4732-4735 (1999).

\bibitem{moon04b}
H.~S. Moon, L.~Lee, K.~Kim, and J.~B. Kim, \enquote{Laser frequency stabilizations using electromagnetically induced transparency,} Appl. Phys.
  Lett. \textbf{84}, 3001 (2004).

\bibitem{yoon04}
T.~H. Yoon, C.~Y. Park, and S.~J. Park, \enquote{Laser-induced birefringence in a wavelength-mismatched cascade system of inhomogeneously broadened Yb atoms,} Phys. Rev. A \textbf{70}, 061803 (2004).

\bibitem{bellsc07}
S.~C. Bell, D.~M. Heywood, J.~D. White, J.~D. Close, and R.~E. Scholten,
  \enquote{Laser frequency offset locking using electromagnetically induced transparency,} Appl. Phys. Lett. \textbf{90}, 171120 (2007).

\bibitem{mullerdt08}
S.~M\"{u}ller, \enquote{Nonlinear optics involving Rydberg states in a rubidium vapor cell,} Diploma Thesis (2008), Universit\"{at} Stuttgart, unpublished.

\bibitem{abel08}
R.~P. Abel, A.~K. Mohapatra, M.~G. Bason, J.~D. Pritchard, K.~J. Weatherill, U.~Raitzsch, and C.~S. Adams,
\enquote{Laser frequency stabilization to highly excited state transitions using electromagnetically induced transparency in a cascade system,}
Appl. Phys. Lett. \textbf{94}, 071107 (2009).

\bibitem{corwin98}
K.~L. Corwin, Z.~T. Lu, C.~F. Hand, R.~J. Epstein, and C.~E. Wieman,
\enquote{Frequency-stabilized diode laser with the Zeeman shift in an atomic vapor,} Appl. Opt. \textbf{37}, 3295-3298 (1998).

\bibitem{robins02}
N.~P. Robins, B.~J.~J. Slagmolen, D.~A. Shaddock, J.~D. Close, and M.~B. Gray,
  \enquote{Interferometric, modulation-free laser stabilization,} Opt. Lett.
  \textbf{27}, 1905-1907 (2002).

\bibitem{wasik02}
G.~Wasik, W.~Gawlik, J.~Zachorowski, and W.~Zawadzki,
\enquote{Laser frequency stabilization by Doppler-free magnetic dichroism,} Appl. Phys. B.
  \textbf{75}, 613-619 (2002).

\bibitem{maric05}
M.~Maric and A.~Luiten,
\enquote{Power-insensitive side locking for laser frequency stabilization,} Opt. Lett. \textbf{30}, 1153-1155 (2005).

\bibitem{gea95}
J.~Gea-Banacloche, Y.-Q. Li, S.-Z. Jin, and M.~Xiao,
  \enquote{Electromagnetically induced transparency in ladder-type inhomogeneously broadened media: Theory and experiment,} Phys. Rev. A
  \textbf{51}, 576-584 (1995).

\bibitem{okoshi80}
T.~Okoshi, K.~Kikuchi, and A.~Nakayama,
\enquote{Novel method for high resolution measurement of laser output spectrum,} Electron. Lett.
  \textbf{16}, 630-631 (1980).

\end{thebibliography}





\end{document}